\documentclass{aa}
\usepackage{amssymb}
\usepackage{graphicx}
\usepackage{hyperref}

\def\asec{\ifmmode ^{\prime\prime}\else$^{\prime\prime}$\fi}

\def\degs{\ifmmode ^{\circ}\else$^{\circ}$\fi}
\def\amin{\ifmmode ^{\prime}\else$^{\prime}$\fi}
\def\asec{\ifmmode ^{\prime\prime}\else$^{\prime\prime}$\fi}
\def\farcs{\hbox{$.\!\!^{\prime\prime}$}}  

\def\degs{\ifmmode ^{\circ}\else$^{\circ}$\fi}
\def\amin{\ifmmode ^{\prime}\else$^{\prime}$\fi}

\def\cm3{\rm ~cm^{-3}}

\def\psr{PSR~B0656+14}

\def\lsim{\!\!\!\phantom{\le}\smash{\buildrel{}\over
  {\lower2.5dd\hbox{$\buildrel{\lower2dd\hbox{$\displaystyle<$}}\over
                               \sim$}}}\,\,}
\def\gsim{\!\!\!\phantom{\ge}\smash{\buildrel{}\over
  {\lower2.5dd\hbox{$\buildrel{\lower2dd\hbox{$\displaystyle>$}}\over
                               \sim$}}}\,\,}

\begin{document}

\title{The near-UV pulse profile and spectrum of the pulsar \psr\thanks{Based on observations with the NASA/ESA 
{\it Hubble Space Telescope}, obtained at the Space Telescope Science Institute, which is operated 
by the Association of Universities for Research in Astronomy, Inc. under NASA contract NAS5-26555.}}

\author 
{Yu.~A.~Shibanov\inst{1}
\and J.~Sollerman\inst{2}
\and P.~Lundqvist\inst{2}
\and T.~Gull\inst{3}
\and D.~Lindler\inst{3,4}
}

\institute{
Ioffe Physico-Technical Institute, Politekhnicheskaya 26,
St. Petersburg, 194021, Russia
\and
Stockholm Observatory, AlbaNova Science Center, Department
of Astronomy, SE-106 91 Stockholm, Sweden
\and 
Goddard Space Flight Center, Exploration of the Universe Division, Code 667,  USA 
\and 
Advanced Computer Concepts, Inc. Potomac, MD,  USA 
}

\date{Received 04, 03, 2005; accepted 01, 06, 2005}

\titlerunning{PSR~B0656+14 in the near-UV}
\authorrunning{Shibanov et al.}
\offprints{shib@astro.ioffe.ru}

\abstract{
We have observed the middle-aged pulsar 
PSR B0656+14 with the prism and the NUV MAMA detector 
of the Space Telescope 
Imaging Spectrograph (STIS) to measure the pulsar spectrum and 
periodic pulsations in 
the near-ultraviolet (NUV).
The pulsations are clearly detected, 
double-peaked 
and very similar to the optical pulse profile. 
The NUV pulsed fraction is $70\pm12$\%. 
The spectral slope of the dereddened phase-integrated spectrum in the 
$\sim1800 - 3200$~\AA~range is $\sim\alpha_{\nu}=0.35\pm0.5$ which 
together with the high pulse fraction
indicates a non-thermal origin for the NUV emission. 
The total flux in the range $\sim1700-3400$~\AA~is estimated to be 
$3.4\pm0.3\times10^{-15}$erg~s$^{-1}$~cm$^{-2}$ when corrected for 
$E(B-V)=0.03$~mag. At a distance of 288~pc this corresponds to  
a luminosity 
$L_{\rm NUV}=3.4\times10^{28}$~erg~s$^{-1}$ assuming isotropy of the 
emission.
We compare the NUV pulse profile with observations from radio to gamma-rays. 
The first NUV sub-pulse is in phase with the
gamma-ray pulse marginally detected with EGRET, 
while the second NUV sub-pulse
is similar both in shape and in phase with the non-thermal pulse in hard
X-rays. This indicates a single origin of the non-thermal emission in the
optical-NUV and in the X-rays. This is also supported by the
observed NUV spectral slope, which is 
compatible with a blackbody plus power-law fit
extended from the X-ray range,  
but dominated by the power-law component in most of the NUV range.
}

\maketitle

\section{Introduction}
The middle-aged pulsar PSR~B0656+14~(PSR J0659+14~\\14) 
was first discovered in the radio
by Manchester et al.~(1978). The pulsar period, 
$\approx$0.385~s, spin-down 
age, $\sim$1.1$\times 10^5$~years, and several other pulsar parameters 
are given in Table~1. 
This is one of the brightest isolated neutron stars 
(NSs) in the X-ray sky (C\'ordova et al.~1989), 
and also the optical counterpart
is brighter ($V\sim25$) than most 
other middle-aged pulsars (Caraveo et al. 1994). 
This made it possible to study the 
optical pulsations of \psr\
(Shearer et al. 1997; Kern et al. 2003), which has previously only
been convincingly done for a few 
much younger pulsars.
        
Owing to its relative  brightness and proximity, PSR~
B~\\0656+14 is one 
of the most intensively studied isolated NSs in different 
wavelength bands. The combined phase integrated spectrum of the pulsar, 
from radio to gamma-rays (cf. Koptsevich et al. 2001), shows that 
its radiation consists of two parts. One component is non-thermal emission from 
the pulsar magnetosphere described by a power-law spectrum with 
different spectral indices in different spectral ranges. The other component is thermal 
blackbody like emission from the whole surface of the cooling NS and 
from much smaller and significantly hotter polar caps of the pulsar. 

While the thermal emission from the whole surface dominates 
in the extreme UV (EUV) and soft X-rays (e.g., Edelstein 2000; Greiveldinger et al. 1996; 
Zavlin \& Pavlov 2004; De Luca et al. 2005), the 
emission at longer wavelengths
is mainly of non-thermal origin  and has a  negative  spectral slope  
(Pavlov et al. 1996; Pavlov et al. 1997; Koptsevich et al. 2001; Komarova et al. 2003). 
Ground-based and HST UV to near-IR photometry suggest that a
main spectral change  
from the near-IR  
to EUV          
occurs somewhere between the B band and the far-UV (FUV).

\begin{table*}[t]
\caption{
Parameters of \psr\ (from Taylor et al. 1993).}
\label{t:656_prop}
\begin{tabular}{cccccccccccc}
\hline\hline
\multicolumn{6}{c}{Observed}&&\multicolumn{5}{c}{Derived} 
\\
\cline{1-7}\cline{9-12}
$P^a$        & $\dot P$   & $D\!M$       & $l$   & $b$ & $\mu_\alpha^b$
 & $\mu_\delta^b$  && $\tau$ & $B$                   & $\dot E$        
      & $d^b$     
      \\
(ms)     & ($10^{-14})$ & (cm$^{-3}$ pc) & (\degs)   & (\degs) & (mas yr$^{-1}$)   & (mas yr$^{-1}$)   && (Myr)    & (G)     & (erg s$^{-1}$)    & (pc)       \\
\cline{1-7}\cline{9-12}
384.9     & 5.50       & 14.0         & 201.1 & 8.3 & $44.07\pm0.63$        & $-2.40\pm0.29$  && 0.11   & $4.7
\times 10^{12}$  & $3.8 \times 10^{34}$  & $288(^{+33}_{-27})$ 
\\
\hline
\end{tabular}
\begin{tabular}{ll}
$^a$Pulsar period on MJD 51687.0 from Kern et al. (2003). & \\
$^b$Pulsar proper motion and parallax based distance from~\cite{brisk03}. & \\
\end{tabular}
\label{t:basic}
\end{table*}

Here we report on time-resolved HST STIS NUV-MAMA prism spectroscopy of \psr\
in the near-UV (NUV). This is the first time the pulse profile of \psr\
has been revealed in the UV. 
The detection of UV pulsations from another
middle-aged pulsar, Geminga,  and the younger
Vela-pulsar was
very recently reported by Kargaltsev et al.
(2005) and Romani et al.~(2005), but was previously only 
obtained for the Crab pulsar (NUV, Gull et al.~1998; FUV, 
Sollerman et al. 2000). 
In this paper we   
study the properties of the
NUV emission of \psr\ and compare them with
available data in other wavelength ranges.
We describe our
observations and  
data reduction in Sect. 2. 
The results are discussed 
in Sect. 3,  and we  
then draw our conclusions in Sect. 4.
\begin{figure}[tbh]
\begin{center}
\includegraphics[width=86mm,  
clip]{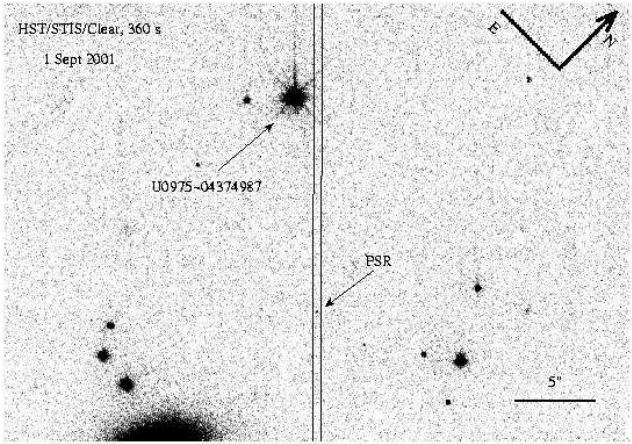}
\end{center}
\caption{
A portion of the PSR B0656+14 field obtained with the 
HST/STIS CCD using the clear filter mode. The long thin lines indicate 
the boundaries of the  52\arcsec$\times$0\farcs5\ aperture for the first visit with 
PA=$-$135\degr.3. The pulsar is indicated as the faint, point-like object 
centered in the aperture.
We have also marked the
reference star, U0975-04374987, used to center the aperture on the 
pulsar position.      
} \label{stis-ima}
\end{figure}
\section{Observations and data analysis}
\subsection{Observations}
Time-resolved spectroscopy of \psr\  in the near-UV ($\sim1700 - 3400$~\AA )
range was performed with the 
HST/STIS during two visits in fall 2001 using the UV
PRISM, the 52\arcsec$\times$0\farcs5\  slit 
and the 
NUV MAMA detector in TIMETAG recording mode. 
The PRISM mode was used because of the very low pulsar flux. 
The relative time resolution was 125$\mu$s. 
Accurate target coordinates were obtained from earlier WFPC2 
images.
The target acquisition was performed using a B=17.9 field star, 
U0975-04374987, which 
is positioned 13\farcs9\ away from the pulsar 
(Fig.~1).
We requested 
and received two visits,  
with the first being on 1 
September and the second on 16 November. 
The slit position angle was PA=$-135$\degr.3 east of north
during the first visit and $-122$\degr.45 during the second visit. 
These position angles allowed us to avoid
any contamination of the pulsar flux by a faint extended background 
object only $\sim$1\arcsec\ north of our target
(Koptsevich et al.~2001; Komarova et al.~2003) 
as well as to exclude other nearby objects 
(Fig.~1).
The two visits also 
allowed initial data evaluation and checks to ensure proper astrometry 
had been used.
Details of the individual science observations are listed in 
Table~\ref{t:log}. 
\begin{table}[tbh]
\caption{Log of observations}
\label{t:log}
\begin{tabular}{lcl}
\hline
\hline
Start Time        & Exp. Time       &         Mode    \\
(MJD)             &    (s)          &                 \\ 
\hline
\hline
52153+ &   Visit 1: 01 September   & 2001   \\   
 \hline
 \hline  
0.9207267  &        120.0$\times$3           &
 STIS   CCD$^a$      \\  
0.9406226	& 1060.0		& NUV
MAMA$^b$ \\
0.9866874	& 2900.2		& NUV MAMA \\
1.0540379       & 2830.2		&NUV  MAMA \\
\hline
\hline
52229+   &  Visit 2: 16 November   &2001 \\   		         
\hline
\hline
0.4670000 &	120.0   &	STIS CCD$^c$ \\
0.4736552 &	1800.0  & NUV MAMA \\
0.5277756 & 2900.2 & NUV MAMA \\
0.5945344  &	120.0   &	STIS CCD$^c$ \\
0.6011900 &	2310.2	&NUV MAMA \\
0.6614096 &	2900.2	&NUV MAMA \\
0.7287130 &	2850.2	&NUV MAMA \\
\hline
\end{tabular}
\begin{tabular}{l}
$^a~$The exposures were obtained to provide
the composite\\ 
  image in Fig.~1.\\ 
$^b~$All NUV-MAMA exposures were done using the PRISM \\
and the 52\arcsec$\times$0\farcs5\ aperture in the
TIMETAG mode.\\
$^c~$Exposures taken through the  52\arcsec$\times$0\farcs5\ 
aperture.\\
\end{tabular}
\end{table}

The initial visit, three orbits in length, began with normal 
acquisition plus three STIS/CCD clear aperture images,
each with 120~s exposure time, for a control image of the pulsar position. 
Figure~\ref{stis-ima} shows the composite field image 
with the 52\arcsec$\times$0\farcs5\  aperture boundaries inscribed. 
An internal 
lamp-illuminated image of the 52\arcsec$\times$0\farcs5\ slit was 
then recorded, and then 
the pulsar was viewed through the same 
aperture by the CCD to confirm that the pulsar 
was  well-centered in the slit.
This also precisely determines the pulsar position for the prism dispersion 
wavelength reference. The remainder of the first visit was allocated to 
three science exposures with the PRISM and NUV MAMA detector (Table~2). 
Internal WAVECAL exposures  
were interspersed between science exposures 
to ensure good reference for wavelength scale.
The total UV PRISM science exposure time in the first visit 
was 6790 s.

The second visit was five orbits long.
CCD 
images of the pulsar viewed through the 
slit were performed 
at the beginning of the first and the third orbits to check 
for any target or image drift internal to the telescope,
but no such drift was detected.  
Five TIMETAG exposures, 
one during each orbit, were recorded 
during 12760~s (Table~\ref{t:log}).
The total TIMETAG science exposure time 
for the two visits was thus 19.55 ks.  
Additional information can be obtained through the HST archive\footnote{archive.stsci.edu} under 
GO program 9156\footnote{presto.stsci.edu/observing/phase2-public/9156.pro}.
\subsection{Data reduction}
 \subsubsection{Extraction of the Spectrum} 
All science exposures for each visit were co-added.  The spectrum was then 
extracted and calibrated using the STIS CALSTIS
pipeline tasks with a 7 pixel wide extraction slit 
(0\farcs013/pix). The backgrounds were extracted 
with a 7 pixels wide slit centered 20 pixels above and below the center of 
the pulsar spectrum. The associated wavecals recorded between science 
exposures were used to adjust the zeropoint of the wavelength scale. 
The two extracted spectra were then registered and averaged with 
weights scaled to the total exposure times for each visit. As the PRISM 
dispersion increases with decreasing wavelength and the detector 
sensitivity falls, we binned the flux below 2000 \AA\ in 4 pixel bins.
Between 
2000 \AA\ and 2300 \AA, the binning was for 2 pixels and above 2300 \AA, 
the data were left at the original PRISM resolution.
This resulted in bins from $\sim20$~\AA\ in the blue to $\sim60$~\AA\ towards the red.
The detector sensitivity and prism dispersion cut the sensitivity 
considerably below 2000 \AA, but in the range 
$2000 - 3200$~\AA~the pulsar is detected 
with S/N $\ga3$ per bin.  
The NUV MAMA detector  
is the dominating source of the background and the
total sky+detector background
was $\sim$700 and
$\sim$3600 counts for the 
1st and 2nd visits, while the signal+background was
$\sim$1200 and $\sim$5000 counts,
respectively.  
 \subsubsection{Generation of the Pulse Profiles}
The pulse profile analysis was done with STIS
Instrument Development Team
tools designed 
originally for testing the TIMETAG mode of the MAMA detectors  (Gull et 
al. 1998). These tools are very similar in function to the STSDAS/STIS  
routines. Both sets of software have been used on the Crab pulsar data 
by Gull et al. (1998) and by Sollerman et al. (2000) and provide very 
similar results. The pulse profile was determined for each visit by 
concatenating the time-tagged events for each PRISM science exposure. 
Event times were corrected for the Earth's  motion around the Sun
and for HST's motion around the Earth using 
the definitive orbital ephemeris extracted from the HST archive.
The orbital position was tabulated at 
one minute intervals and was 
determined by cubic-spline interpolation for each event.

The pulse period was determined from the data stream by the following 
process. All events within a rectangular patch centered on the pulsar 
spectrum were selected. Likewise, we selected all events in background patches 
above and below the pulsar spectrum.
The pulse period was determined by 
maximizing the RMS of the pulse profile of the target events 
(see Sollerman et al. 2000). 
For the two visits, the computed periods are:
\newline	Visit 1: 0.384902 s
\newline	Visit 2: 0.384903 s
\newline 
This is in excellent agreement to within the last digit with 
the radio pulsar period (Table~\ref{t:basic}).
Using the 
ephemeris from Kern et al. (2003) we obtain:
\newline	Visit 1: 0.3849028 s
\newline	Visit 2: 0.3849031 s
\newline 
at the epochs of our observations.
The measaured  period is also in agreement
with earlier radio observations presented by \cite{Taylor} 
and \cite{chang99}, and there is thus no evidence for glitches in \psr\ 
over
the last 16 years.
This also ensures us that the obtained pulse profile is correct.
To increase the 
S/N, the data were binned into phase  bins at intervals 1/16th that of 
the pulse period. The background data were likewise binned, but showed 
no structure with phase. We then subtracted 
the combined 
background average bin count. 
The TIME-TAG mode provided relative time resolution of 125~$\mu$s, 
but absolute time determination is not known to better 
than one second due to design limitations of the STIS computer timing 
updates. 
Internal clock stability is sufficient to ensure good
time stability from orbit to orbit, even up to several days apart, but not for a ten week interval. We were thus forced 
to use the pulse profile shapes in the NUV and in the optical to phase align the two visits and to determine the NUV 
pulse position with respect to the radio pulse (see Sect.~3.1 and 3.2).
\section{Results and Discussion}
\subsection{The NUV pulse profile}
\begin{figure}[tbh] 
\begin{center}
\includegraphics[width=85mm,  clip]{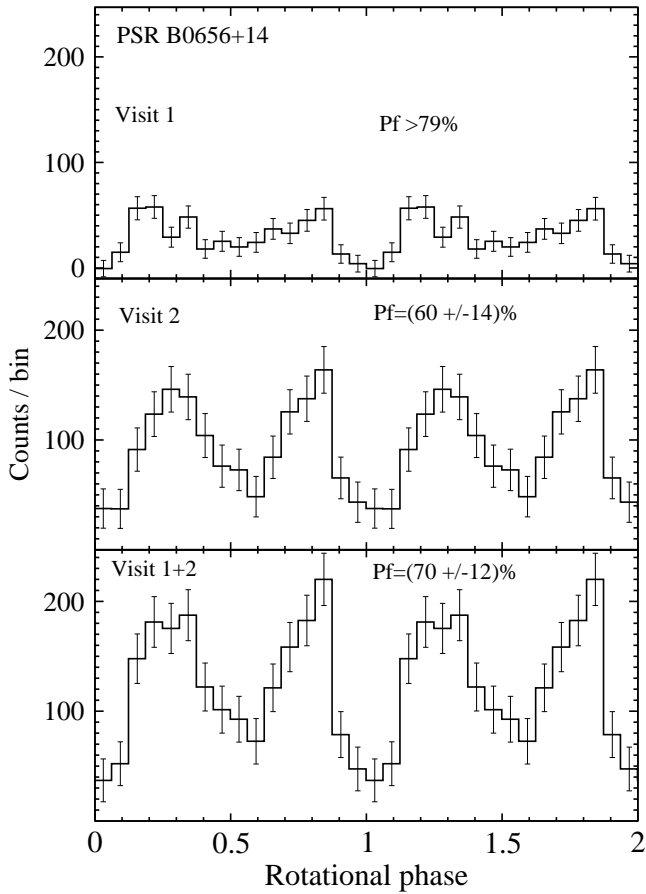}
\end{center}
\caption{The 
background subtracted pulse profiles of  PSR B0656+14 in the near-UV 
from visit 1 (top), visit 2 (middle), and after combining of the data from the two
visits (bottom).  
} \label{f:nuv-profiles}
\end{figure}

The background subtracted spectral integrated pulse
profiles obtained from the data of visits 1 and 2
are shown in the {\sl top} and {\sl middle} 
panels of Fig.~\ref{f:nuv-profiles}, respectively.    
While the pulsations in the first
visit are only significant at the $\sim4\sigma$ level, the
significance of the 
pulsations in the second, longer visit 
is $\sim8\sigma$.
The pulse profile contains two sub-pulses whose
maxima are separated by $\sim0.5$ in phase 
and are connected by an emission ``bridge''. The shape of
the first sub-pulse is rather symmetric,
while the second sub-pulse is characterized by a linear
flux increase in its leading part as well as by a sharp
edge in the trailing part. This sharp drop
and the bridge can also be seen in the first visit
profile. 
We have used 
these features
to align both
pulse profiles in phase and combined
them.
The result is shown in the {\sl bottom} panel of  
Fig.~\ref{f:nuv-profiles}. The significance of the
pulsations increased in the combined profile to
$9.2\sigma$. 

We cannot exclude the presence of a persistent (or a
weakly variable) flux component from the pulsar in the
NUV range. As shown in Fig.~2, the estimated pulse
fraction Pf (the ratio of the number of counts above the
minimum of the profile to the total number of counts) is in
the range $58\% - 82\%$~representing 
a $1\sigma$ confidence interval 
based on count statistics.
The persistent flux 
contribution into the pulse integrated flux is
thus within the   $18 - 42\% $  range.  
The lower boundary of this 
uncertainty range is compatible with the published
upper limit $\la16$\%
in the optical range (Kern et al. 2003).
\subsection{Pulse profiles from the radio to gamma-rays} 
The shape of the observed NUV double pulse profile
is very similar to what is seen in the optical range
by Kern et al.~(2003). This supports a single emission
mechanism being responsible for the emission in both 
ranges. 
We note that the Kern et al.~(2003) pulse profile was 
actually very different from the
optical pulse profile for \psr\ earlier reported by 
Shearer et al.~(1997). 
Kern et al.~(2003) 
discussed whether this large discrepancy could be due to differences in
pass bands, or due to a  pulse profile 
variation with time. Our NUV profile clearly 
supports the findings of Kern et al. (2003) and disfavours the 
explanations to accomodate the earlier profile by 
Shearer et al. (1997).

\begin{figure}[t] 
\begin{center} 
\includegraphics[ 
width=64.6mm, 
bb=29 178 401 
1042, clip]{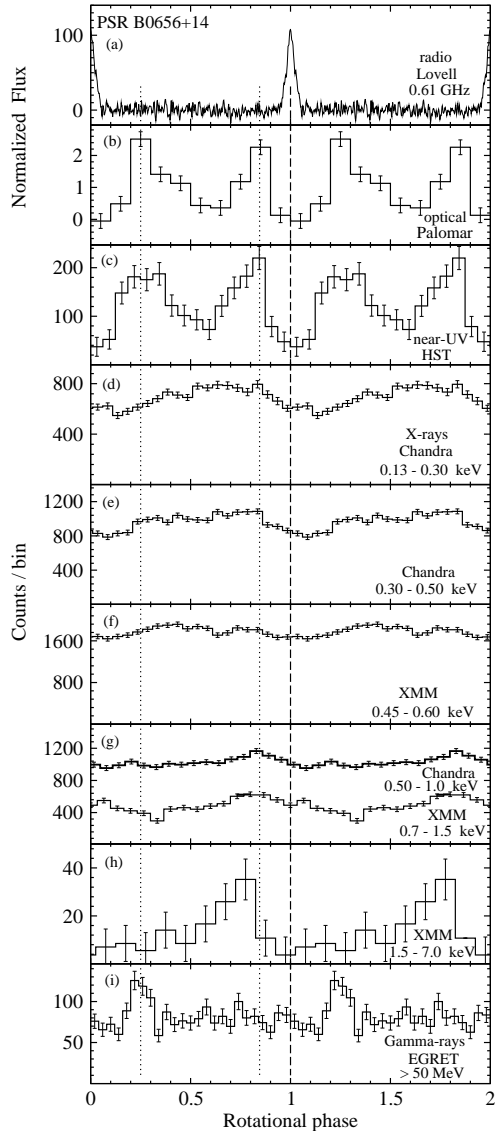} 
\end{center} 
\caption{
Pulse profiles of \psr\ measured by different
telescopes from radio through gamma-rays. Dashed and 
dotted lines mark the position of the radio and the two optical 
pulses, respectively. 
Note the similar morphology of the optical and NUV 
profiles and the hints of the 
soft X-ray counterparts to the optical sub-pulses seen 
roughly at the same phases in the soft
X-rays. The second NUV sub-pulse is similar in
shape and position to the 
pulse in hard X-rays, 
while the first NUV sub-pulse coincides 
with the gamma-ray pulse.
The dynamic range for the various 
pulse profiles are chosen to illustrate the 
different pulse fractions in different spectral
bands. 
To better see the actual pulse shapes in the soft X-ray range, 
we refer to Fig.~\ref{HST-Chandra}, as well as 
to Pavlov et al.~(2002), Zavlin \&
Pavlov~(2004), and  De Luca et al.~(2005).
}
\label{f:nuv-profiles2} 
\end{figure}

The optical pulse profile by Kern et al. (2003)  
was obtained with absolute timing 
which allowed a phase alignment between 
the optical and the radio pulse profiles. 
The same was done for X-ray data obtained with the Chandra and   
XMM-Newton observatories
(Pavlov et al.~2002; Zavlin \& Pavlov
2004; De Luca et al.~2005), and for gamma-ray data obtained with EGRET  
(Ramanamurthy et al.~1996). We have used the similarity of the optical and
NUV profiles to align our data with the radio pulse
and to compare  
its absolute position in phase and
morphology  with the available data
in a wide spectral range, from radio
through gamma-rays
(Fig.~\ref{f:nuv-profiles2}). 

In the radio (Gould \& Lyne 1998)
the pulsar has a sharp single pulse with a
width of about 0.2 of the NS rotational
period 
(Fig.~3a).
In the soft
X-rays it instead displays shallow (Pf$\sim10-24$\%) 
pulsations with a sine-like light curve. It has a
single broad maximum per pulsar period, typical 
for the thermal emission from the 
surface of a NS with non-uniform temperature distribution over the
surface 
(Fig.~3d).
This is completely 
different from the optical and  NUV profiles 
(Figs.~3b--c)
and underline
the difference in the emission mechanisms  
dominating 
in the optical-NUV (non-thermal) and in the soft X-rays (thermal).
However, in the smooth light curves in soft X-rays one can
discover small, but significant, features
approximately coinciding in phase with the positions of
the NUV sub-pulses 
(Figs.~3c and 3d--g).
This can be more easily seen in the zoomed 
examples of the X-ray profiles (Fig.~4, see also Pavlov et
al.~2002 and  Zavlin \& Pavlov 2004).
With increasing photon energy, where the thermal emission is
assumed to be dominated 
by the pulsars hot polar caps, the
maximum of the X-ray profile is shifted
toward the position of the 2nd NUV
sub-pulse  
(Fig.~3g).
However,  
this
emission can not contribute significantly 
to the NUV 
because the hot spot area is too small
($R_{HS}\la 2$~km, De Luca et al. 2005). 
In the hard X-rays
(Fig.~3h),
where the
spectrum is dominated by a non-thermal magnetospheric
component, both the position and  
the morphology of the profile is 
similar to that of the 2nd NUV sub-pulse.  The counterpart of the 1st NUV sub-pulse is  
not observed at the low S/N available   
in the hard X-ray range,  while it coincides in position with  
the gamma-ray pulse marginally detected
with the EGRET (Ramanamurthy et al. 1996).
The pulsar has not yet been reliably detected
in the intermediate energy range between the hard X-rays 
and gamma-rays.     
Available RXTE observations obtained in the 15-250 keV range
provide only an upper limit on the pulsar flux
and a very uncertain pulse profile   
(Chang \& Ho 1999). 

To summarize,  
the observed profiles suggest that the
pulsed emission outside the radio range is a sum of
at least two components. One component, likely
non-thermal emission, displays a two peak pulse profile which is
clearly resolved in the optical and in the NUV. The other
component, likely thermal emission, has a shallow broad
sine-like profile with a single pulse per period. 
This component could potentially be partially responsible for
the persistent and bridge emission seen in the NUV, 
since its maximum 
coincides with  
the NUV inter-pulse position. 
The counterparts of the optical-NUV sub-pulses can be
resolved from the thermal background in soft X-rays.
The 2nd  NUV sub-pulse counterpart is clearly 
resolved  in hard X-rays, while the 1st
sub-pulse may be seen in gamma-rays. 
The coincidence of the position and morphology 
of the 2nd NUV sub-pulse and 
the X-ray profile supports a strong 
correlation of the non-thermal emission in these 
ranges, which has also been deduced from     
phase averaged spectral analysis 
(e.g., Koptsevich et al. 2001).
The radio pulse is centered at the off-pulse 
emission in all other ranges.   
\subsubsection{Thermal nature of the bridge and persistent emission?}
The variation of the thermal flux with the NS rotation may 
be understood in terms of anisotropic heat conduction 
through the strongly magnetized envelope of a
cooling NS. 
The regions of the star with radial  magnetic
fields are hotter than the regions with  tangential fields 
(e.g., Greenstein \& Hartke 1983; 
Page 1995; Shibanov \& Yakovlev 1996; Potekhin \& 
Yakovlev 2001). 
The observed flux modulation 
in the soft X-rays 
is in agreement with 
a $\sim$14\%\ 
modulation 
expected from a middle-aged cooling NS with 
a dipole magnetic field of a few times 
$10^{12}$ G
(e.g., Potekhin \& Yakovlev 2001). 
A single maximum per period with a relatively symmetric pulse profile  
suggests that only one magnetic pole is visible in
the X-rays during 
the pulsar rotation. This is also compatible with   the
pulsar viewing  geometry derived from the radio (Everett \& Weisberg 2001)
and optical (Kern et al. 2003) polarization measurements.
However, the $\sim0.5$ phase lag between the maxima
in the radio and  soft X-rays suggests 
that the hollow polar cap magnetospheric cone assumed to
be responsible for the radio and optical emission    
must be strongly curved with height. 
Alternatively,
the magnetic field surface structure is different from a simple
centered dipole configuration.  The latter
is supported  by the significant phase lag  (De Luca
et al.~2005)  between 
the pulse maximum of the cool 
blackbody  component  from the whole
surface of the NS and the hot blackbody 
component from the hot spots  
(Figs.~3d and 3g).

To investigate if the NUV bridge  
flux is dominated by the thermal emission we 
have applied a simplified model (Greenstein \& Hartke  1983)  assuming 
that the NUV and soft
X-rays fluxes ($F$) vary with rotational phase ($\Phi$) as 
$F=F_{ \parallel}cos^2(\pi \Phi+\alpha) +F_{\perp}sin^2(\pi \Phi+\alpha) $, 
where  $F_{ \parallel}$,  $ F_{\perp}$ are the fluxes at the minimum
and maximum of the  pulse profile, respectively. These are assumed to 
originate from  the   surface regions  with predominantly 
radial and tangential magnetic 
fields, and $\alpha$ accounts for a small shift ($\Delta \Phi\approx 0.1$)  
of the profile  minimum from  the zero phase.            
\begin{figure}[tbh]
\begin{center}
\includegraphics[width=85mm,  clip]{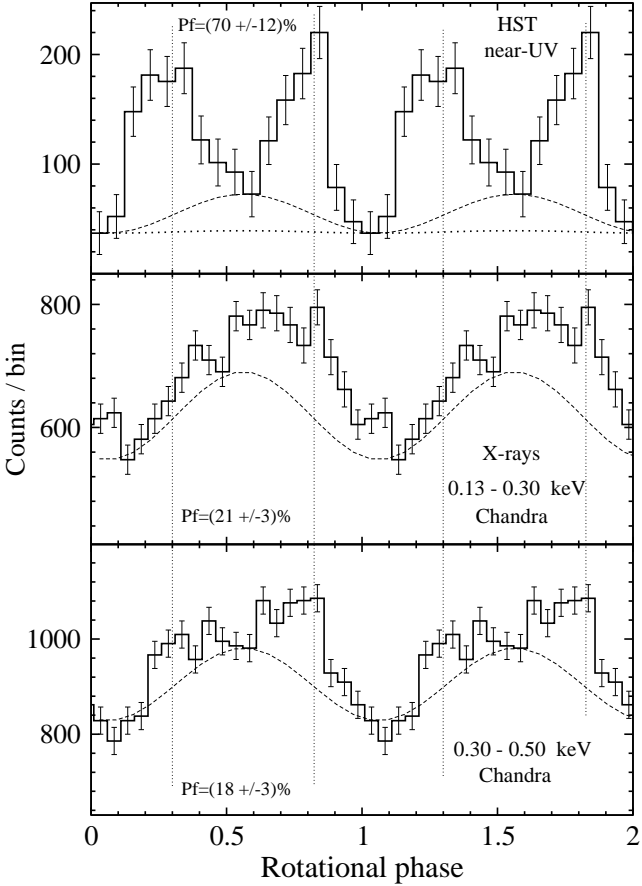}
\end{center}
\caption{Enlarged NUV and soft X-ray  pulse
profiles. The dashed lines are model profiles  of  thermal
emission from a NS with  anisotropic surface temperature distribution 
(e.g., Greenstein \& Hartke  1983). 
The minimum and maximum fluxes of the
model are specific for  each  energy range and 
are selected  to be equal to the count levels  around the phases 
corresponding to  the minimum and the inter-pulse bridge  of the
NUV-profile, respectively. The dotted horizontal line in the upper panel 
represents  the  
extrapolation of the X-ray model profile of the thermal emission to the
NUV (see text for details).   
} \label{HST-Chandra}
\end{figure}
 
 Figure~\ref{HST-Chandra} shows that this  model  
provides a qualitative fit (dashed curves) to the 
soft X-ray profiles and to the smooth part of the NUV 
profile including the bridge. 
 It also brings out  the non-thermal  counterparts of the NUV sub-pulses 
 in  X-rays. 
However, assuming a blackbody 
 spectrum of the thermal emission  with a maximum in
 the soft X-rays (e.g., Zavlin \& Pavlov~2004),  
 and  that  the temperature $ T  \propto F^{1/4}$ in this range, we 
  can estimate  from  the X-ray  model light curves  with  
  $ F_{ \parallel}/F_{\perp} \approx 1.25 $    
 that the  effective temperature variations averaged over the visible 
 surface of the NS are $\sim5$\%.  
The smooth flux variation over the  NUV
is larger:  $ F_{ \parallel}/F_{\perp}\approx 2$.
 Taking into account that any thermal NUV is in the Rayleigh-Jeans (RJ) part 
 of the blackbody spectrum where      
 $T  \propto F $,  we obtain much larger temperature variations 
 ($\ga$50\%) from the NUV fit. This 
 contradiction with the X-rays  can  
 hardly be accounted  for by any 
 variations of the visible  emitting area
 during the NS rotation ($\la
 10\%$, De  Luca et al.~2005).
An extrapolation of  
the estimated  X-ray temperature variations to the NUV (Fig.~4, dotted
 horisontal curve), shows that the pulsed 
bridge emission in the NUV can not be due to this
mechanism, but the resulting NUV emission is 
compatible with the observed persistent flux.  
This means that the bridge emission is probably not
thermal. It could instead result from overlapping of 
the two non-thermal sub-pulses. 

To further investigate if the NUV persistent flux
is thermal, we can compare its relative
contribution to the total flux, $\lesssim 30$\%,  with the expected
 contribution from the thermal  cool
 blackbody component provided by extrapolation  
 of  the combined two blackbody
 $+$ power-law fit obtained in X-rays for
 the phase averaged spectrum  (e.g.,
 Koptsevich et al.~2001) to
 the NUV.
Such a model is further discussed in Sect.~3.3 and shown in Fig.~5, and
yields $\sim 25$\%\ thermal contribution at 
 2800~\AA, where the PRISM throughput peaks.
 This is consistent with the unpulsed fraction in the NUV observations.
 For comparison, 
 the expected  blackbody contribution  in
 the optical range at 5000~\AA\ 
is several times smaller,
 $\sim 7$\%. In this region, 
Kern et al.~(2003) were not able 
to detect any persistent flux at $\la16$\%.
In this interpretation, we are actually
resolving the surface of the NS during the off-pulse in the NUV.
Future observations, 
allowing for phase resolved spectroscopy in
the NUV, FUV, and  optical as well as  detailed
comparison with the phase resolved
spectroscopy in the X-rays,  are necessary to confirm  
such an interpretation.
\subsection{The near-UV spectrum}
The combined phase averaged spectrum of
the pulsar is shown in   Fig.~5. 
The spectrum is rather noisy at the edges of the
observed range, while in the range $\sim2150 - 3200$~\AA, where the
NUV-MAMA/PRISM has its maximal throughput, the S/N is $\ga3$ per bin.
The spectral energy distribution is rather flat with
the flux
apparently increasing
with increasing frequency.
A power-law (PL; $F_{\nu} \propto \nu^{\alpha_{\nu}}$)
fit to the spectrum results in a spectral index of
$\alpha_{\nu}=0.1\pm0.6$. This is indicated by the flat dotted line in 
Fig.~5.
The slope may be slightly positive, but a negative index is
also consistent with the data.

To check the absolute flux calibration we compared the
NUV spectrum with the available broadband NUV
photometric data obtained with the HST/FOC in the
F342W and F195W bands  (Pavlov et al. 1997).
Within the uncertainties the
spectral data are consistent with the photometric magnitudes.
\begin{figure}[tbh]
\includegraphics[width=86mm, clip]{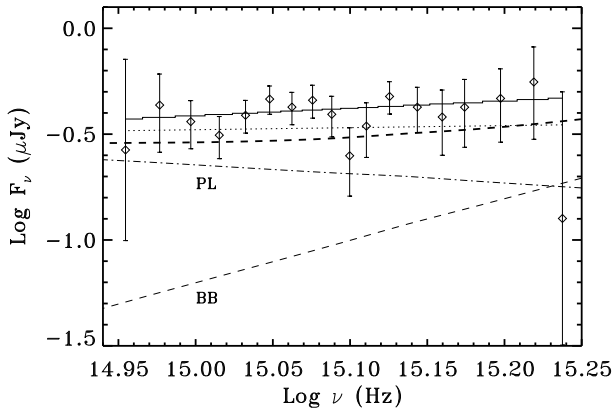}
\caption{
The dereddened spectrum of  PSR B0656+14 using $E(B-V)=0.03$.
The data have been binned.
The best power-law fit is shown by the solid line. It has a
spectral index of
$\alpha_{\nu}=0.35\pm0.5$.
To demonstrate that the effect of the reddening is small we also show the
best power-law fit for the non-dereddened data (dotted line).
We have also included the 
contributions from a cool blackbody
(BB, dashed), power-law (PL, dash-dot), and their sum
(BB+PL; thick dashed)   resulting from the spectral fit
of X-ray data and extended to the NUV (Koptsevich et al. 2001).
} \label{f:nuv-spec3}
\end{figure}

The interstellar color excess $E(B-V)$ toward PSR B0656+14  has been
estimated to be in the range of $0.01 - 0.05$ mag with a
likely value of $E(B-V)=0.03$ mag
(Pavlov et al. 1996; Kurt et al.~1999).
The dereddened [$E(B-V)=0.03$] spectral index is
$\alpha_{\nu}=0.35\pm0.5$. 
This is shown by the solid line in Fig.~5.
Dereddening the spectrum by this amount gives an integrated flux in
the range $\sim1700-3400$~\AA~of
$3.4\pm0.3\times10^{-15}$erg~s$^{-1}$~cm$^{-2}$.
At a distance of 288~pc this corresponds to the
 luminosity  
$L_{\rm NUV}=3.4\times10^{28}$~erg~s$^{-1}$, 
assuming  isotropy of the emission.
Even for an extinction of $E(B-V)=0.05$ mag, the one sigma limit of the
spectral index  remains smaller than unity. This is significantly lower 
than
the index of the RJ part of the blackbody spectrum ($\alpha_{\nu}=2$),
and strongly favors a non-thermal
component for the bulk of the emission in the NUV range.
To state this in a different way, to reach the RJ spectral index would 
require 
an extinction of $E(B-V)=0.2$, which is much outside the suggested range.
Even the maps of Schlegel et al. (1998) only indicate $E(B-V)=0.09$
throughout the entire Galaxy along this line-of-sight.

The observed NUV spectrum is actually compatible
with the extension of the absorbed combined spectral
fit, two blackbodies (BB) $+$ power-law (PL),
of the X-ray data (the hot blackbody component
does not contribute significantly in the
NUV and is not shown in Fig.~5).
Although such an extrapolation is uncertain (see, e.g., Kargaltsev et al. 2005), we note that the 
shape of the dereddened sum BB+PL (Fig.~5, thick dashed line) 
in the NUV is very similar to the 
observed NUV spectrum, 
although the PL clearly dominates in most of the NUV 
range. 
A non-thermal origin for the bulk of the NUV emission is also consistent
with our analysis of the pulse profile.
The change-over from the PL-component to the BB-slope 
seems to occur at the high frequency boundary
of the observed NUV-range (Fig.~5).
If this is true, we expect
a higher spectral slope and a larger fraction of 
the persistent flux in the FUV range.

A possible absorption feature
is hinted at $\sim$2400~\AA\  (15.10 in Log~$\nu$).
The formal significance is low 
($\lesssim2\sigma$, see Fig.~5), but a flux
depression around this wavelength is present in both visits.
If confirmed by future observations, such a feature could be interpreted 
as an electron/positron cyclotron line formed at
$B\approx 5\times 10^8$~G.
Assuming the  dipole surface field
obtained from the  pulsar spin-down rate
(Table~1) this places   the absorbing
plasma at 220 km above the NS surface.
This
is roughly consistent with the minimum
height ($\ga350$~km) for the optical emission
estimated from the assumed critical synchrotron
frequency
by Kern et al.~(2003).
The faintness of the
absorption feature suggests that
the obscuring magnetospheric plasma
is optically thin and cannot  affect
the radiation in X-rays. 
No spectral features or depressions
have been detected in the soft X-rays.
This disfavors interpretations
of the phase lags between the maxima of
the cool and hot BB profiles and the
radio pulse based on existence of a  magnetospheric ``blanket"
affecting the pulse shape (e.g., De Luca et al.~2005 and references
therein).  
These phase lags could instead be the
result of a strong inhomogeneity of the
surface magnetic field of the NS. 
On the other hand, the spectral energy
distribution  of  PSR B0656+14 in the
optical/near-IR range appears to be non-monotonic and shows 
potential depressions 
(Koptsevich et al.~2001; Komarova et al.~2003). 
These features are  deeper than
the possible absorption line in the NUV. This could be
explained by increasing  the optical
depth  of the absorbing magnetospheric plasma 
with the wavelength.  Optical spectroscopy and
timing observations of the pulsar in different
optical bands are needed to study the nature of
the dips and their relation to the marginal
absorption line in the NUV.               
\section{Conclusions}
We have detected the near-UV pulsations of
PSR B0656~\\+14 using the STIS NUV-MAMA onboard the HST.
The pulse period we derive is in agreement with previous estimates.
The pulse profile is double-peaked with an inter-pulse bridge and is very
similar to the profile detected in the optical range by Kern et al. (2003). 
Sharp pulses and a high NUV pulse fraction, $70\pm12$\%,
suggest a non-thermal origin of the
emission.
  
We have compared the NUV pulse profile with pulse profiles from the 
radio to gamma-rays.
The first NUV sub-pulse is in phase with the
marginal gamma-ray pulse, while the second NUV sub-pulse is
similar in both shape and phase 
with the pulse detected in the hard X-rays.
This favors a single origin
for the non-thermal pulsed part of the emission
from the optical to the X-ray range, as
previously has also  been seen from
the broadband photometric observations 
in the optical and NIR   
(e.g., Koptsevich et al. 2001).
The NUV sub-pulse counterparts can also be resolved from the
shallow soft X-ray profiles dominated by 
thermal emission from the surface of the NS.

A simple model for the thermal X-ray flux, in terms of an anisotropic
heat distribution of the NS surface, does not explain the pulsating
NUV bridge emission, which is therefore likely non-thermal. 
However, any persistent flux in the NUV could contribute 
at the 30$\pm12$\% level, which is potentially
higher than the upper limit $\la 16$\%
measured in the optical (Kern et al. 2003). 
This is in qualitative agreement with the thermal flux from the 
abovementioned model, 
and we therefore argue that the non-pulsed NUV emission is
due to the RJ part of the thermal emission component.

We have also measured the NUV spectrum of  PSR B0656+14.
The phase averaged spectrum is basically flat,
$\alpha_{\nu}=0.35\pm0.5$.  
This slope is consistent with contributions from both a thermal
component  
from the whole surface of the cooling NS and from a magnetospheric
power-law  
component extrapolated from the X-rays. 
The main spectral change from the power-law 
dominating magnetospheric emission to 
the thermal RJ from the NS surface is likely to occur 
at the  boundary between the NUV and FUV ranges. 
Preliminary reports on FUV emission from PSR B0656+14 
indicate that the  
FUV is indeed thermal (Pavlov et al. 2004). 
This also agrees with our model where the non-pulsed NUV flux is
thermal. 
We therefore predict a higher spectral slope and a  
lower pulse fraction 
in the FUV. 

We note that a similar situation applies to the middle-\\-aged 
Geminga pulsar (Kargaltsev et al. 2005, their Figs. 7 and 10). 
However,  the younger Vela pulsar (Shibanov
et al. 2003; Romani et al. 2005) and the Crab pulsar 
(Sollerman 2003) display a relatively flat
spectrum from the near-IR to UV range.
Moreover, the non-thermal emission of the young Crab pulsar and the Crab twin
PSR B0540-69 (Boyd et al. 1995) display pulse-profiles
that also does not change significantly from the optical 
to the UV. The same seems to be true for Geminga
 (Kargaltsev et al.~2005)  and for Vela  (Romani et al.~2005).
This could indicate that a unique mechanism, which apparently does not 
strongly depends on pulsar age, drives the non-thermal pulsed emission in the optical and UV.

\begin{acknowledgements}
We are grateful for help from Alexei  
Kop-\\-tsevich in the initial phase of this study.
Part of this research has made use of the database of published pulse 
profiles maintained 
by the European Pulsar Network.
This work was supported by NASA and The Royal Swedish Academy of Sciences.
YAS were supported by the RFBR (grants
03-02-17423, 03-07-90200 and 05-02-16245) and
RLSS programme 1115.2003.2.
The research of PL is further sponsored by the Swedish Research Council.
PL is a Research Fellow at the Royal Swedish Academy supported by a grant 
from the Wallenberg Foundation. 
DL was supported in part by funding through GO proposal 9156. 
TG and DL were supported in part through the
STIS Guaranteed Time Observations resources.
\end{acknowledgements}

{}
\clearpage

\end{document}